\documentclass[12pt]{iopart}
\usepackage{iopams} 
\usepackage{graphicx} 
\begin{document}

\title{Modelling incomplete fusion dynamics of weakly-bound nuclei at near-barrier energies}

\author{Alexis Diaz-Torres}

\address{Department of Physics, Faculty of Engineering and
Physical Sciences, University of Surrey, Guildford, Surrey GU2 7XH,
United Kingdom}
\ead{a.diaztorres@surrey.ac.uk}
\begin{abstract}
The classical dynamical model for reactions induced by weakly-bound nuclei at near-barrier energies is developed further. It allows a quantitative study of the role and importance of incomplete fusion dynamics in asymptotic observables, such as the population of high-spin states in reaction products as well as the angular distribution of direct alpha-production. Model calculations indicate that incomplete fusion is an effective mechanism for populating high-spin states, and its contribution to the direct alpha production yield diminishes with decreasing energy towards the Coulomb barrier. It also becomes notably separated in angles from the contribution of no-capture breakup events. This should facilitate the experimental disentanglement of these competing reaction processes. 
\end{abstract}


\section{Introduction}
Nuclear physics research has entered a new era with developments of radioactive nuclear beam 
facilities, where nuclear reactions are the primary probe of the new physics, such as novel structural changes. In those facilities, the low-energy nuclear reactions research is highly focused on understanding astrophysically important reaction rates involving exotic nuclei. These are often weakly-bound with a few-body, cluster structure that can easily be dissociated in their interaction with other nuclei. Understanding the breakup mechanism and its impact on nuclear reaction dynamics is essential. A major consequence of breakup is that a rich scenario of reaction pathways arises, such as events where (i) not all the resulting breakup fragments might be captured by the target, termed incomplete fusion ({\sc icf}), (ii) the entire projectile is captured by the target, called complete fusion ({\sc cf}), and (iii) none of the breakup fragments are captured, termed no-capture breakup ({\sc ncbu}). 

Since the availability of intense exotic beams is still limited, extensive experimental research has recently been carried out exploiting intense beams of stable weakly-bound nuclei, such as $^{6,7}$Li and $^{9}$Be \cite{Nanda1,Beck0}. Understanding the effect of their breakup on near-barrier fusion has been a key aspect of these investigations \cite{Canto1}. These have definitively demonstrated that breakup suppresses the above-barrier fusion cross sections. Most recently, experimental activities have been focused on disentangling breakup and competing reaction mechanisms from inclusive and exclusive coincidence measurements \cite{Signorini1,Shrivastava,Beck1,Santra1,Souza1}. One of the challenges is to obtain a complete quantitative understanding of the breakup mechanism and its relationship with near-barrier fusion. This research is guided by complete sub-barrier breakup measurements \cite{Ramin1}.  

Theoretical works have addressed the low-energy reaction dynamics of weakly-bound nuclei using quantum mechanical, classical and mixed quantum-classical approaches \cite{Jeff1,CDCC1,TCSM1,CDCC2,Yabana,Hagino,CDCC3,CDCC4}. Among these, the continuum-discretised coupled channels ({\sc cdcc}) framework has been very successful \cite{Jeff1,CDCC2,CDCC3,CDCC4}. However, existing quantum models have limitations \cite{Ian1}, as they cannot calculate integrated {\sc icf} and {\sc cf} cross sections unambiguously. Neither, after the formation of {\sc icf} products, can they follow the evolution of the surviving breakup fragment(s) since {\sc icf} results in depletion of the total few-body wave function. 

These difficulties are overcome by the three-dimensional classical dynamical reaction model suggested in Ref. \cite{Alexis1}. A crucial \emph{input} of this model is a stochastically sampled breakup function proposed in Ref. \cite{Hinde1}, which can be determined from sub-barrier breakup measurements \cite{Ramin1,Hinde1}. This function encodes the effects of the Coulomb and nuclear interactions that cause the projectile breakup. Hence, this approach is \emph{not} a breakup model, rather it is a quantitative dynamical model for relating the sub-barrier {\sc ncbu} to the above-barrier {\sc icf} and {\sc cf} of weakly-bound nuclei \cite{Hinde1}. The model has been successfully applied to interpreting fusion measurements of weakly-bound nuclei \cite{Santra1,Ramin1}, and isomer ratio measurements \cite{Gasques1}. 

In this paper, I report on further developments of this classical dynamical model. The key new aspect is the time propagation of the surviving breakup fragment and the {\sc icf} product, allowing the description of their asymptotic angular distribution and dynamical variables. These should be very useful in (i) current experimental activities aimed at disentangling competing reaction mechanisms from asymptotic observables such as alpha-production yields \cite{Souza1}, (ii) interpreting particle-$\gamma$-coincidence experiments \cite{Singh}, and (iii) applications to $\gamma$ ray spectroscopy \cite{Dracoulis1,Mullins1,AndreaJungclaus}. The new developments are illustrated with a simplified test case that does \emph{not} aim at adjusting any measurements. It is worth mentioning that various dynamical models have been proposed for multi-fragmentation and {\sc icf} in heavy-ion induced reactions at energies well-above the Coulomb barrier 
($ \gtrsim 10$ MeV/nucleon), as recently reviewed in Ref. \cite{Singh}. In this different context, some of those models (e.g., Refs. \cite{Moehring,Bondorf}) use concepts and techniques (e.g., classical trajectory, Monte Carlo sampling) that overlap with some involved in the present model. The model is explained in Section 2. In Section 3, numerical results are discussed, and a summary is given 
in Section 4.   

\section{Model}
 
The main features of the model are as follows: 
\begin{description}
\item \textnormal{(i)} The target $T$ is initially at rest in the origin of the laboratory frame, whilst the weakly-bound (two-body) projectile $P$ approaches the target (along the z-axis) with incident energy $E_0$ and orbital angular momentum $L_0$. For each $L_0$ (chosen to be an integer number of $\hbar$) an \emph{ensemble} of $N$ incident projectiles is considered. Including the $P-T$ mutual Coulomb and nuclear forces, classical equations of motion determine an orbit with a definite distance of closest approach $R_{min}(E_0,L_0)$.

\item \textnormal{(ii)} The complexity of the projectile dissociation is empirically encoded in a density of (local breakup) probability ${\cal P}^L_{BU} (R)$, a function of the projectile-target separation $R$, such that ${\cal P}^L_{BU}(R) dR$ is the probability of breakup in the interval $R$ to $R+dR$ (see Appendix A). A key feature is that for a given projectile-target combination, both measurements \cite{Ramin1,Hinde1} and {\sc cdcc} calculations \cite{Alexis1} indicate that the integral
of this breakup probability density along a given classical orbit is an
exponential function of its distance of closest approach,
$R_{min}(E_0,L_0)$: 
\begin{equation}
{P}_{BU}(R_{min})= 2\int_{R_{min}}^\infty {\mathcal P}^L_{BU}(R) dR 
= A\,\exp (-\alpha R_{min}). \label{one}
\end{equation}
Consequently, ${\cal P}^L_{BU} (R)$ has the same exponential form, ${\mathcal P}^L_{BU}(R) \propto \exp (-\alpha R)$. (The factor of 2 highlights that breakup may occur along the entrance or exit branch of the trajectory, although the exponential form will clearly place maximum probability of breakup at $R_{min}$.) This function is sampled to determine the position of breakup in the orbit discussed in (i). In this position, the projectile is instantaneously broken up into fragments F1 and F2. These interact with $T$, and with each other, through real central two-body potentials having Coulomb barriers $V_{B}^{ij}$ at separations $R_{B}^{ij}$, $i,j=1,2,T,\ i\neq j$.
 
\item \textnormal{(iii)} The instantaneous dynamical variables of the excited projectile at breakup, namely its total internal energy $\varepsilon_{12}$, 
its angular momentum $\vec{\ell}_{12}$ and the separation of the fragments
$\vec{d}_{12}$ are all Monte Carlo sampled. 
The initial separation $d_{12}$ between the fragments and its orientation 
$\vec{d}_{12}$ are determined by sampling the radial and angular probability 
distributions of the projectile ground-state (g.s.) wave function, respectively. 
For a two-body projectile with $0^{+}$ g.s., a very good approximation for calculating $d_{12}$ is through a Gaussian sampling function in the classically allowed region of the fragments, whilst the orientation of $\vec{d}_{12}$ is isotropic. (This will be used in the test case below.) For high ${\ell}_{12}$ excitations, when there is no barrier between F1 and F2, $d_{12}$ is equated with their external turning point.
The orientation of $\vec{\ell}_{12}$ is chosen randomly from all directions orthogonal to
$\vec{d}_{12}$.  ${\ell}_{12}$ is 
sampled uniformly in the interval [0,$\ell_{max}$], whilst for $\varepsilon_{12}$ an exponentially decreasing function for energies between the 
top of the barrier ($V_{B}^{12}$) and a chosen maximum $\varepsilon_{max}$ is sampled. Both $\ell_{max}$ and $\varepsilon_{max}$ are increased until 
convergence of the observables occur.

\item \textnormal{(iv)} Having fixed the position and dynamical variables of the excited projectile 
fragments at the moment of breakup, the instantaneous velocity of the particles
F1, F2 and $T$ is determined by conservation of energy, linear
momentum and angular momentum in the overall center-of-mass frame 
(see Appendix B). 
These breakup initial conditions are transformed to the laboratory frame where the three bodies are propagated in time. The calculated trajectories of F1, F2 and $T$ determine the number of {\sc icf}, {\sc cf} and {\sc ncbu} events, fragment F$j$ being assumed to be captured if the classical trajectories take it within the fragment-target barrier radius $R_{B}^{jT}$.

\item \textnormal{(v)} From the $N$ breakup events sampled for each projectile 
angular momentum $L_0$, the numbers of events $N_i$ in which $i=$ 0
({\sc ncbu}), 1 ({\sc icf}), or 2 ({\sc cf}) fragments are captured
determine the relative yields $\widetilde{P_i}=N_i/N$ of these three
reaction processes after breakup, with $\widetilde{P_{0}} +
\widetilde{P_{1}} + \widetilde{P_{2}} = 1$. The absolute probabilities
$P_i(E_0,L_0)$ of these processes are expressed in terms of the 
relative yields and the integrated breakup probability over the whole 
trajectory $P_{BU}(R_{min}$):
\begin{eqnarray}
P_{0}(E_0,L_0) &=& P_{BU}(R_{min})\,\widetilde{P_{0}},\label{eq5} \\
P_{1}(E_0,L_0) &=& P_{BU}(R_{min})\,\widetilde{P_{1}},\label{eq3} \\
P_{2}(E_0,L_0) &=& [1-P_{BU}(R_{min})]\,H(L_{cr}-L_0) \nonumber\\
               &+& P_{BU}(R_{min})\, \widetilde{P_{2}},
\label{eq1}
\end{eqnarray}
where $H(x)$ is the Heaviside step function and $L_{cr}$ is the critical partial wave for projectile fusion. The cross sections are calculated using
\begin{equation}
\sigma_i (E_0)= \pi \lambda ^2 \sum_{L_0} (2L_0 + 1) P_i(E_0,L_0),
\label{eq2}
\end{equation}
where $\lambda ^2=\hbar^2/[2m_P E_0]$ and $m_P$ is the projectile mass. 

Beside the absolute cross sections (\ref{eq2}), asymptotic observables, 
such as the angle, kinetic energy and
relative energy distributions of the fragments from {\sc ncbu} 
events, are calculated by tracking their trajectories to a large 
distance from the target. 

\item \textnormal{(vi)} For the {\sc icf} events, the time propagation of the {\sc icf} product and the surviving breakup fragment is now incorporated into this model as follows. The fragment F$j$ (after overcoming the Coulomb barrier $V_{B}^{jT}$) reaches the target radius, forming the {\sc icf} product, while the other fragment flies away. At this moment, the three-body propagation turns into a two-body propagation, with definite interaction potentials and initial conditions. These are determined by the position and velocity of the three particles, at the moment when the {\sc icf} product is formed. These also yield the spin and excitation energy distributions of the \emph{primary} {\sc icf} product. The asymptotic angular distribution of the {\sc icf} product and the surviving breakup fragment is calculated in terms of their trajectories.

\end{description}

\section{Numerical results}
This model is implemented in the {\sc platypus} code, as described in 
Ref. \cite{Alexis2}. In order to illustrate the new developments of (vi), calculations are carried out at a laboratory energy of $E_0$ = 45 and 65 MeV, for the reaction of a pseudo-$^8$Be projectile $P$ (modeled as a weakly-bound $s$-state of two $\alpha$-particles \cite{Alexis1}) with the $^{208}$Pb target $T$. 

The breakup function ${P}_{BU}(R_{min})$ in eq. (\ref{one}) has parameters 
$A=5.98\times 10^3$ and $\alpha=0.85$ fm$^{-1}$, which was deduced in Ref. \cite{Alexis1} from mapping fusion measurements for the $^9$Be + $^{208}$Pb system with this test reaction. This is because the model is limited to solving a three-body problem. However, very recent measurements \cite{Ramin1} have shown that prompt $^9$Be breakup occurs dominantly through an excited $^8$Be nucleus, validating the approximation of a $^9$Be projectile by $^8$Be.

The nuclear interaction between the alpha particle and the {\sc icf} product $^{212}$Po is considered to be the Woods-Saxon potential ($V$, $r$, $a$) $\equiv$
(33.98 MeV, 1.48 fm, 0.63 fm) resulted from the global Broglia-Winther parametrization \cite{Broglia1}. (Please note that in the potential the radius parameter is multiplied by $A_T ^{1/3}$.) The rest of the model parameters are the same as in Ref. \cite{Alexis1}.

\begin{figure}
\begin{center}
\includegraphics[width=0.50\textwidth,angle=0]{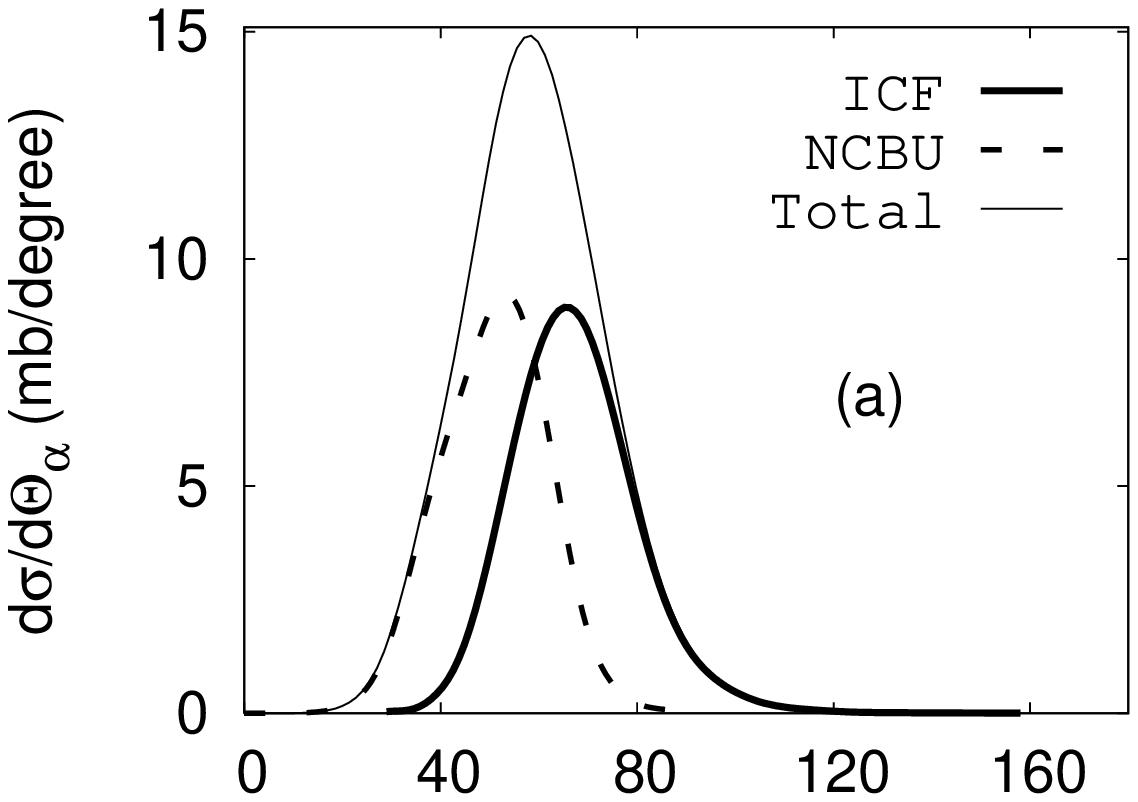} \\
\includegraphics[width=0.50\textwidth,angle=0]{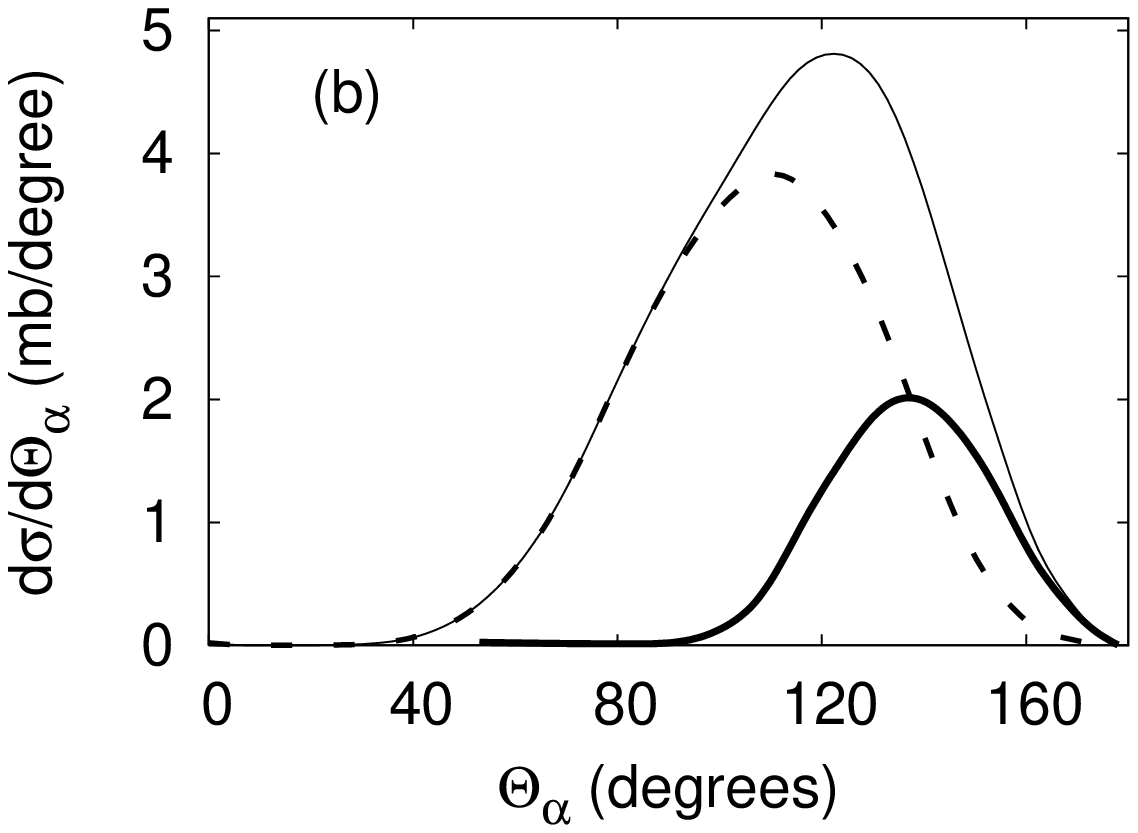}
\caption{Angular distribution of direct alpha-production 
for $^8$Be + $^{208}$Pb at two laboratory energies $E_0$: 
(a) 65 MeV, and (b) 45 MeV. 
With decreasing energy towards the Coulomb barrier, 
the {\sc ncbu} events dominate, separating its 
centroid substantially from that of the {\sc icf} events. 
The total alpha-production distribution changes its shape notably.} 
\label{Figure1}
\end{center}
\end{figure}

\begin{figure}
\begin{center}
\includegraphics[width=0.50\textwidth,angle=0]{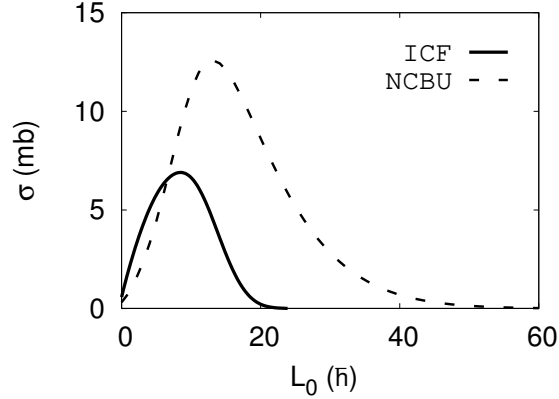}
\caption{{\sc icf} and {\sc ncbu} cross sections as a function of 
the relative angular momenta $L_0$ for $^8$Be + $^{208}$Pb 
at $E_0 = 45$ MeV. The contribution of high-partial waves 
shifts the {\sc ncbu} distributions in Fig. 
\ref{Figure1} to smaller angles, with respect to the 
{\sc icf} distributions.} 
\label{Figure2}
\end{center}
\end{figure}

Figure \ref{Figure1} shows the angular distribution of 
direct alpha-production for two laboratory energies near the $P-T$ s-wave Coulomb barrier ($39.9$ MeV), namely (a) $E_0 = 65$, and (b) $45$ MeV. The contribution of the {\sc icf} and {\sc ncbu} events is represented by thick solid and thick dashed lines, respectively. Their incoherent sum is represented by the thin solid line. Its shape significantly changes as the incident energy decreases. While the contribution of the {\sc icf} and {\sc ncbu} events appears to be similar at well-above barrier energies [panel (a)], the {\sc ncbu} contribution gradually dominates with decreasing energy towards the barrier [panel (b)]. Here, its centroid significantly separates from the centroid of the {\sc icf} contribution. Both centroids shift to higher angles as the incident energy decreases, due to the reduction of relative partial waves affecting these reaction processes. However, the {\sc ncbu} centroid always remains lower than the {\sc icf} centroid, as expected. This is because higher partial waves contribute to the {\sc ncbu} process (see Fig. \ref{Figure2}). Fig. \ref{Figure2} presents the incident angular momentum distribution of the {\sc icf} (solid line) and {\sc ncbu} (dashed line) processes at a laboratory energy of $E_0 = 45$ MeV.

\begin{figure}
\begin{center}
\includegraphics[width=0.50\textwidth,angle=0]{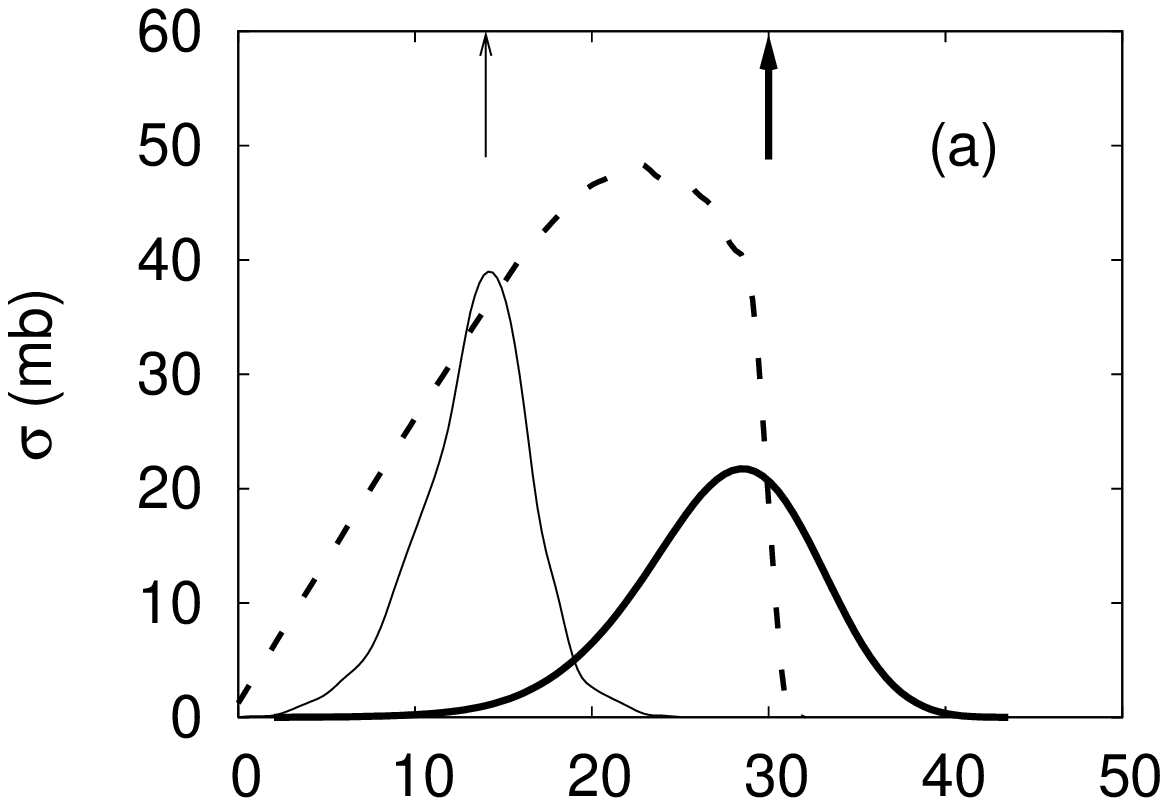} \\
\includegraphics[width=0.50\textwidth,angle=0]{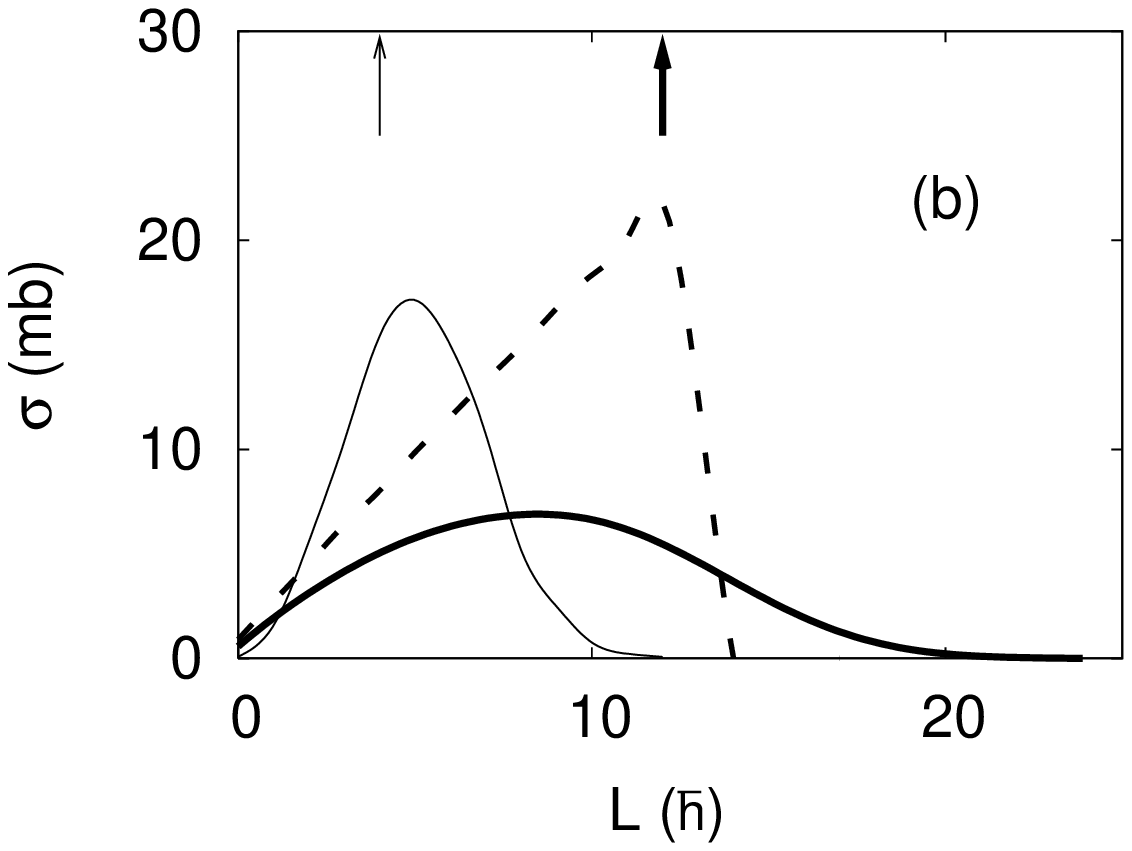}
\caption{Spin distribution of \emph{primary} fusion products resulting from the $^8$Be + $^{208}$Pb reaction at two laboratory energies $E_0$: (a) 65 MeV, and (b) 45 MeV. The {\sc cf} ($^{216}$Rn) spin distribution is presented by the dashed line, whilst the {\sc icf} ($^{212}$Po) spin distribution is shown in two representations, \emph{i.e.}, in terms of (i) the $^8$Be incident angular momentum $L_0$ (thick solid line), and (ii) the angular momentum brought in by the $\alpha$-particle into the $^{208}$Pb target (thin solid line). The thick arrow denotes $L_{cr}$ for $^8$Be fusion as an inert projectile, whereas the thin arrow is $l_{cr}$ for fusion of a direct beam of $\alpha$-particles (with half of $E_0$) on $^{208}$Pb. The relevant features are: (1) the {\sc icf} mechanism increases the angular momentum brought in by the $\alpha$-particle into $^{208}$Pb (thin solid line goes beyond the thin arrow), (2) the relative localization of the $L_0$-window for {\sc cf} and {\sc icf} significantly depends on $E_0$ (dashed and thick solid lines), strongly overlapping as $E_0$ decreases, and (3) {\sc cf} following $^8$Be breakup becomes substantial with increasing $E_0$, shifting down the maximum of the {\sc cf} spin distribution (dashed line) with respect to the thick arrow.}       
\label{Figure3}
\end{center}
\end{figure}
 
Figure \ref{Figure3} shows the angular momentum distribution of \emph{primary} {\sc icf} 
($^{212}$Po) and {\sc cf} ($^{216}$Rn) products at (a) $E_0 = 65$, and (b) $45$ MeV. The 
{\sc icf} spin distribution is represented in terms of both the $^{8}$Be incident angular momentum $L_0$ (thick solid line) and the angular momentum brought in by the $\alpha$-particle into the $^{208}$Pb target (thin solid line). These representations result in two very different {\sc icf} angular momentum distributions, both providing relevant features of the {\sc icf} dynamics as explained below. 

Comparing the thick solid line with the {\sc cf} spin distribution (dashed line) it is seen that the localization of the $L_0$-window for {\sc icf} significantly depends on the incident energy $E_0$. Their overlap strongly increases as $E_0$ decreases towards the barrier [panel (b)], indicating that {\sc icf} and {\sc cf} are two competing reaction processes at near-barrier energies. However, {\sc cf} following $^{8}$Be breakup is here a very small component ($1.5\%$) of the {\sc cf} cross section ($149.2$ mb). Instead, either one of the $\alpha$-particles is captured by the target, contributing to the {\sc icf} cross section ($84.7$ mb), or the two $\alpha$-particles survive and contribute to the {\sc ncbu} cross section ($238.8$ mb). At $E_0 = 65$ [panel (a)], {\sc cf} following 
$^{8}$Be breakup is very substantial, representing $41\%$ of the {\sc cf} cross section 
($981.5$ mb). This significantly shifts down the maximum of the {\sc cf} spin distribution (dashed line) with respect to the critical angular momentum $L_{cr}$ for $^{8}$Be fusion as an inert projectile (thick arrow). Here, $L_0$ values around $L_{cr}$ determine the {\sc icf} cross section ($273.6$ mb) which is similar to the {\sc ncbu} cross section ($259.3$ mb).

\begin{figure}
\begin{center}
\includegraphics[width=0.50\textwidth,angle=0]{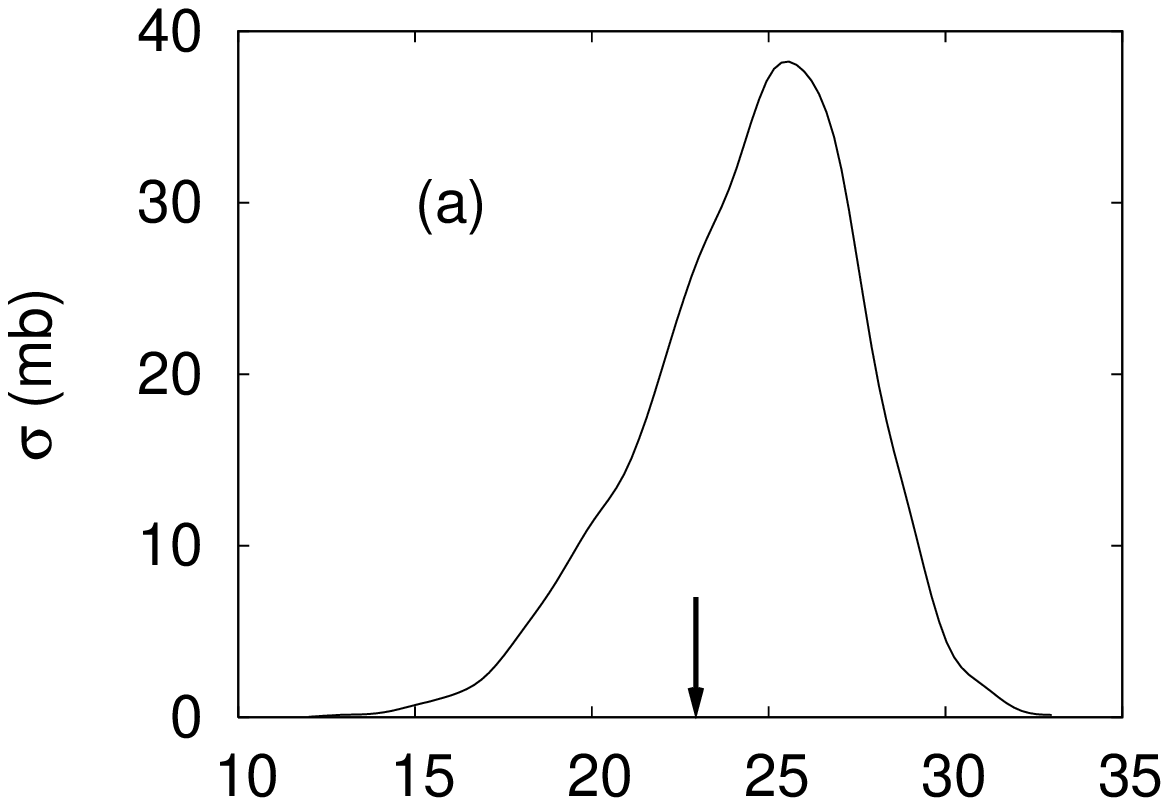} \\
\includegraphics[width=0.50\textwidth,angle=0]{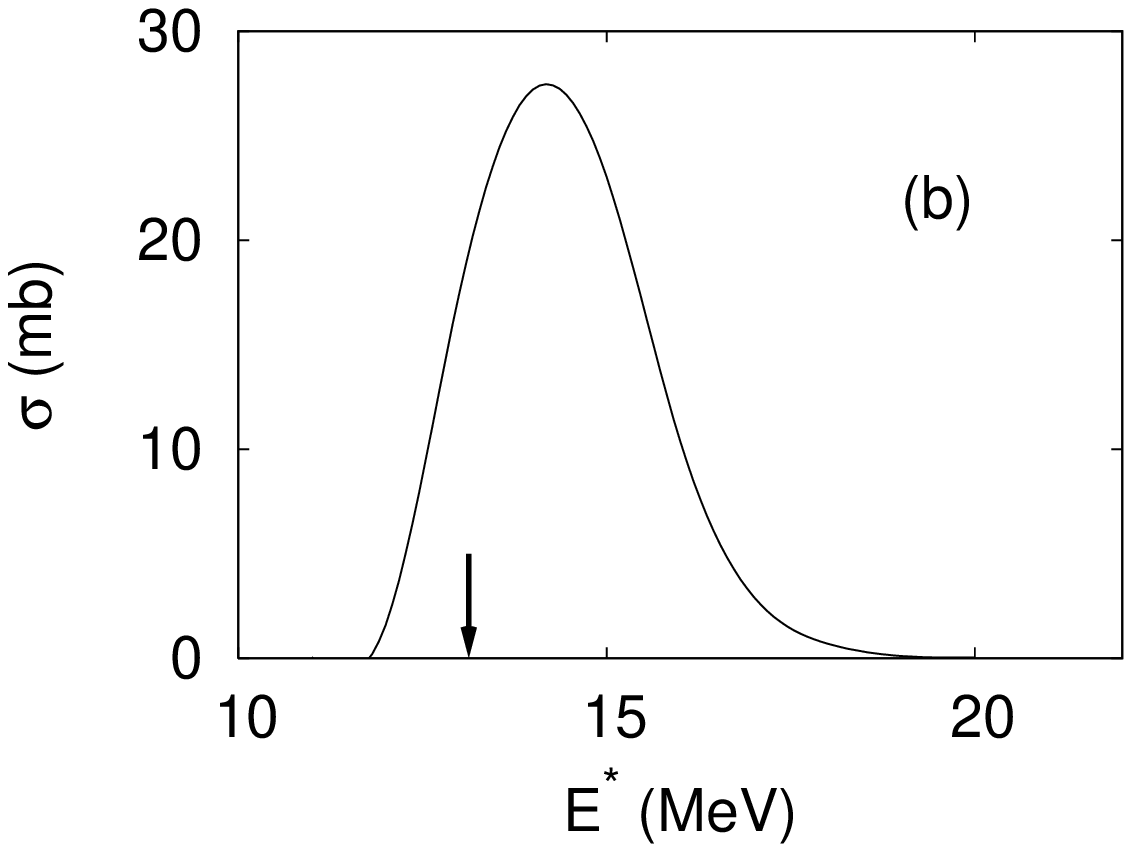}
\caption{Excitation energy distribution of the \emph{primary} {\sc icf} product $^{212}$Po 
resulting from the $^8$Be + $^{208}$Pb reaction at two laboratory energies $E_0$: (a) 65 MeV, and (b) 45 MeV. The arrow denotes the 
excitation energy of $^{212}$Po, when this compound nucleus is formed by fusion of a direct beam of $\alpha$-particles, with half of $E_0$, on $^{208}$Pb. Following breakup of $^8$Be, the fusing $\alpha$-particle is significantly affected by the interaction with the surviving $\alpha$-particle. This spreads the $^{212}$Po excitation energy over a range of values. The distribution becomes narrower and its maximum approaches the arrow, as $E_0$ decreases towards the barrier.} 
\label{Figure4}
\end{center}
\end{figure}

The {\sc icf} spin distributions (thin solid line) in Fig. \ref{Figure3} are the crucial ones in order to access the effectiveness of the {\sc icf} mechanism for populating 
high-spin states in $^{212}$Po. It is observed that the tail of these distributions goes well beyond the thin arrow which denotes $l_{cr}$ for fusion, on the $^{208}$Pb target, of a direct beam of $\alpha$-particles with half of $E_0$. Clearly, the extra torque caused by the interaction between the fusing and surviving $\alpha$-particles enhances the angular momentum of the {\sc icf} product. This interaction also affects the excitation energy of $^{212}$Po substantially, as presented in Fig. \ref{Figure4}.                  

Fig. \ref{Figure4} shows the excitation energy distribution of the \emph{primary} {\sc icf} product $^{212}$Po (thin solid line) at (a) $E_0 = 65$, and (b) $45$ MeV. It is caused by the interaction between the fusing and surviving $\alpha$-particles during the {\sc icf} process. The arrow corresponds to the pre-determined excitation energy of $^{212}$Po, when this nucleus is formed through {\sc cf} of a direct beam of $\alpha$-particles (with half of $E_0$) on $^{208}$Pb. This value remains smaller (by a few MeV) than the excitation energy associated with the maximum of the distribution. The shape of the distribution also changes with $E_0$, becoming narrower as $E_0$ decreases towards the barrier. These distributions along with the {\sc icf} spin distributions in Fig. \ref{Figure3} (thin solid line) are vital for a reliable prediction of the yield of final {\sc icf} products. To my knowledge, these aspects have not been yet included in broadly used evaporation codes which mostly describe {\sc cf} evaporation residues.

The present model can be applied to more realistic cases (e.g., $^{11}$Be  induced reactions), provided all the necessary information for Monte Carlo sampling is known (e.g., the breakup function and the projectile g.s. wave-function). When one of the breakup fragments is neutral, its capture by the target nucleus can be assumed to occur when its trajectory takes it within the target radius. Although not implemented yet, the model can also predict the elastic angular distribution of the projectile. This distribution can easily be calculated in terms of the projectile-target orbits and the probability of the weakly-bound projectile's survival (see Appendix A), i.e., this probability weighting the contribution of the different projectile-target orbits. Direct reaction processes other than the elastic breakup may significantly contribute to observables associated with {\sc ncbu}, {\sc cf} and {\sc icf}. These are not included in the model yet. However, their contributions could be included using transfer and inelastic-breakup functions. Of course, these would make the calculations more complex, as these direct reaction processes result in additional bifurcation points along the projectile-target orbits. Nevertheless, their inclusion as well as the treatment of more complex projectiles (e.g., $^{6}$He and $^{11}$Li) will be interesting, necessary developments of the present approach.   

\section{Summary}

The classical dynamical model for reactions of weakly-bound nuclei at near-barrier energies has been developed further. It allows us to quantify the role and importance of {\sc icf} dynamics in asymptotic reaction observables, such as the angular distribution of the direct alpha-production. The {\sc icf} contribution to this yield diminishes with decreasing energy towards the barrier. However, the {\sc icf} and {\sc ncbu} contributions are clearly separated in angles at near-barrier energies, facilitating the experimental disentanglement of these competing reaction processes. The present developments also provide the spin and excitation energy distributions of primary {\sc icf} products, which are essential ingredients for calculating the yield of final {\sc icf} products with present evaporation codes. The {\sc icf} mechanism appears to be an effective route for producing high-spin states. All these observables may also be affected by other direct processes, such as transfer \cite{Alexis3}, which are not included in the model yet. Nevertheless, the present classical dynamical model is a powerful tool for interpreting fusion measurements involving radioactive nuclei and in applications to $\gamma$ ray spectroscopy. The development of a unified quantum dynamical description of relevant reaction processes ({\sc icf}, {\sc cf}, {\sc ncbu} and transfer) remains a great theoretical challenge. One possibility of tackling this issue could be through a time-dependent density-matrix approach incorporating the concept of quantum decoherence \cite{Ian1,Alexis4}. 

\ack

The author thanks B. Bayman and M. Dasgupta for suggesting Appendix A, and J.A. Tostevin for discussions related to Appendix B. Support from the UK Science and Technology Facilities Council (STFC) Grant No. ST/F012012/1 is acknowledged.

\appendix

\section{Breakup probability function}

Let us define two probabilities: (i) the probability of breakup between $R$ and $R+dR$, 
$\rho(R)dR$ [being $\rho(R)$ a density of probability], and (ii) the probability of the weakly-bound projectile's survival from $\infty$ to $R$, $S(R)$. The survival probability at $R+dR$, $S(R+dR)$, can be written as follows
\begin{equation}
S(R+dR) = S(R) \, [1 - \rho(R)dR]. 
\label{b1}
\end{equation}
    
Expression (\ref{b1}) suggests the following differential equation for the survival probability $S(R)$,
\begin{equation}
\frac{dS(R)}{dR} = - S(R)\,\rho(R), 
\label{b2} 
\end{equation}
whose solution is [$S(\infty) = 1$]:
\begin{equation}
S(R) = \exp(-\int_{\infty}^{R} \rho(R) dR). 
\label{b3}
\end{equation}

From (\ref{b3}), the breakup probability at $R$, $B(R) = 1 - S(R)$. If 
$\int_{\infty}^{R} \rho(R) dR \ll 1$, $B(R)$ can be written as
\begin{equation}
B(R) \approx \int_{\infty}^{R} \rho(R) dR.
\label{b4} 
\end{equation}   

From (\ref{b4}), identifying $\rho(R)$ with ${\mathcal P}^L_{BU}(R)$, we obtain expression 
(\ref{one}) for the breakup probability integrated along a given classical orbit.

\section{Matching prior- and post-breakup stages}

The integrals of motion in the overall center-of-mass ({\sc cm}) system are the total energy $E_{tot}=\frac{m_T}{(m_T + m_P)}E_0$, the total linear momentum $\vec{P}_{tot}=\vec{0}$, and the total angular momentum $\vec{L}_{tot}=m_P b_0 (\vec{v} - \vec{V}_{CM})$ that is orthogonal to the initial reaction plane. $m_T$, $m_P$, $b_0$, $\vec{v}$, and $\vec{V}_{CM}$ are the mass of the target and projectile, the impact parameter between the projectile and the target, the velocity of the incident projectile in the laboratory system and the {\sc cm} velocity, respectively. 

Just after breakup, the two-body projectile is excited to a definite state ($\varepsilon_{12}$, $\vec{\ell}_{12}$ and $\vec{d}_{12}$), as explained in Section 2. The relative vector between $P$ and $T$ ($\vec{R}_{PT}$) is also known. Thus, the separation between the three bodies is known. The modulus of the velocity between $P$ and $T$ ($V_{PT}=P_{PT}/\mu_{PT}$) results from the total energy conservation 
\begin{equation}
E_{tot} = \varepsilon_{12} + U_{1T}(r_{1T}) + U_{2T}(r_{2T}) + P_{PT}^2 /2\mu_{PT}, \label{a1}
\end{equation}
where $U$ is the interaction potential between the target and the breakup fragments. 

The total linear momentum $\vec{P}_{tot} = \vec{p}_T + \vec{p}_1 + \vec{p}_2 = \vec{p}_T + \vec{p}_{P^*}$, where 
$\vec{p}_{P^*}$ is the momentum of the center of mass of excited $P$ relative to the overall {\sc cm}. We need the velocities of $P$ and $T$ relative to the overall {\sc cm} ($\vec{\widetilde{v}_P}$ and $\vec{\widetilde{v}_T}$) to complete the initial conditions for subsequent propagation in time of the three bodies. These velocities are related to each other by the expressions
\begin{equation}
\vec{\widetilde{v}_T} = - \frac{m_P}{m_T}\, \vec{\widetilde{v}_P}, \label{a2}
\end{equation} 
 \begin{equation}
\vec{V}_{PT} = \vec{\widetilde{v}_P} - \vec{\widetilde{v}_T}, \label{a3}
\end{equation}    
where the magnitude of $\vec{V}_{PT}$ is known through expression (\ref{a1}). To know the direction of this velocity the conservation of total angular momentum is applied. 

The total angular momentum $\vec{L}_{tot} = \vec{\ell}_{12} + \vec{L}_{PT}$, so the angular momentum ($\vec{L}_{PT}$) associated with relative motion of $P$ and $T$ about {\sc cm} is known. This vector can be written as
\begin{equation}
\vec{L}_{PT}=m_P \, \vec{R}_{PT} \, \times \, \vec{\widetilde{v}_P}. \label{a4}
\end{equation}  

We now write $\vec{\widetilde{v}_P}$ in terms of radial and transverse components as follows:
\begin{equation}
\vec{\widetilde{v}_P} = \widetilde{v}_P^{(r)} \, \vec{r} \, + 
                        \, \widetilde{v}_P^{(q)} \, \vec{q}, \label{a5}  
\end{equation}
where $\vec{r} = \vec{R}_{PT}/R_{PT}$ and $\vec{q} = \vec{n} \times \vec{r}$, being $\vec{n} = \vec{L}_{PT}/L_{PT}$. The transverse component 
$\widetilde{v}_P^{(q)} = L_{PT}/(m_P R_{PT})$, and for the target 
$\widetilde{v}_T^{(q)} = - L_{PT}/(m_T R_{PT})$. The radial component is obtained using expressions (\ref{a2})-(\ref{a3}) and knowing the transverse component:
\begin{equation}
\widetilde{v}_P^{(r)} = \pm 
\Big\{ V_{PT}^2 - \Big[ \widetilde{v}_P^{(q)} \, \big( 1 + \frac{m_P}{m_T} \big) \Big]^2 \Big\}^{1/2} 
/ \big( 1 + \frac{m_P}{m_T} \big). \label{a6}
\end{equation}

Both positive and negative roots are consistent with the conservation of the integrals of motion. Hence, both roots are uniformly sampled. Finally, the position and velocity vectors of the projectile fragments and the target are transformed to the laboratory system using Galilean transformations.

\end{document}